\documentclass[aps,prl,twocolumn,superscriptaddress,preprintnumbers,showpacs]{revtex4}
\usepackage{graphics}
\def\eq#1\en{\begin{equation} #1 \end{equation}}

\def\eqa#1\ena{\begin{eqnarray} #1 \end{eqnarray}}

\usepackage{graphicx,color}

\begin{document}

\title{The exotic baryon mass spectrum and \\
the {\bf 10}--{\bf 8} and $\overline{\bf 10}$--{\bf 8} mass difference in the Skyrme model}


\author{G. Duplan\v{c}i\'{c}}
\affiliation{Theoretical Physics Division, Rudjer Bo\v skovi\' c Institute, 
Zagreb, Croatia}
\author{J.Trampeti\'{c}}
\affiliation{Theoretical Physics Division, Rudjer Bo\v skovi\' c Institute, 
Zagreb, Croatia}
\affiliation{Theory Division, CERN, CH-1211 Geneva 23, Switzerland}
\affiliation{Theoretische Physik, Universit\"{a}t M\"{u}nchen, Theresienstr. 37, 80333 M\"{u}nchen, Germany}

\date{\today}

\begin{abstract}
The {\bf 8}, {\bf 10}, and $\overline{\bf 10}$ baryon mass spectrum
as a function of the Skyrme charge $e$ and the SU(3)$_f$ symmetry breaking parameters is given in tabular form. 
We also estimate the decuplet--octet and the antidecuplet--octet mass difference.
Comparison with existing literature is given. 
\end{abstract}


\maketitle
Recently we applied the concept of minimal SU(3) extended Skyrme model
to nonleptonic hyperon and $\Omega^-$decays \cite{dppt}
producing reasonable agreement with experiment.
This concept uses only one free parameter, the Skyrme charge $e$,
and flavor symmetry breaking (SB) term, proportional to $\lambda_8$ in the kinetic 
and the mass term.
The main aim of this brief report is the application of the same concept 
in an attempt to predict the baryonic decuplet--octet ($\Delta$) and antidecuplet--octet ($\overline{\Delta}$)
mass difference as well as to evaluate the mass spectrum for octet, decuplet, and 
the recently discovered antidecuplet baryons. 

The experimental discovery \cite{penta1} and the later confirmation \cite{penta2} 
of the exotic, presumably spin 1/2, baryon 
of positive strangeness, $\Theta^+$,
was recently supported by the NA49 Collaboration \cite{penta3} discovery of the exotic isospin 3/2 baryon 
with strangeness -2, $ \Xi^{--}_{3/2}$. In this way, the antidecuplet, and possibly the other states 
of the higher SU(3)$_f$ representation,
moved from pure theory into the real world of particle physics.

The first successful prediction of mass of one member of the $\overline{\bf 10}$ baryons, 
known as penta-quark or $\Theta^+$-baryon, in the framework of the 
Skyrme model was presented in Ref. \cite{MP}.
Later, many authors used different types of quark, chiral soliton, diquark, etc. 
models \cite{dia1,wei1,WK,K,pra3,Glo,Itz,Kar,Bor,Zhu,Ger,Hua,Bij,JW,DP,Ell}, to estimate 
the higher SU(3) representation (${\overline{\bf 10}},\;{\bf 27}, \; {\rm etc.} $) mass spectrum, 
relevant mass differences and other baryon properties. 

In this brief report, like in Ref. \cite{dppt}, we use the minimal SU(3) extension of 
the Skyrme Lagrangian introduced in \cite{wei}:
\begin{eqnarray}
{\cal L} = {\cal L}_{\sigma} + {\cal L}_{Sk} + {\cal L}_{WZ} + {\cal L}_{SB} ,
\label{2a} 
\end{eqnarray}
where  ${\cal L}_{\sigma}$, ${\cal L}_{\rm Sk}$, ${\cal L}_{\rm WZ}$, and ${\cal L}_{\rm SB}$ denote 
the $\sigma$-model, Skyrme, Wess--Zumino and symmetry breaking Lagrangians 
\cite{sky,tho,wess,witt,adk,adk1,gua,yab,desw}, respectively.

For profile function $F(r)$ we use the arctan ansatz \cite{dia,prat2}:
\begin{eqnarray}
F(r) = 2{\rm arctan} \left[ \left({\frac{r_0}{r}}\right)^2 \right].
\label{11}
\end{eqnarray}
Here $r_0$ - the soliton size - is the variational parameter and the second power of $r_0/r$ is determined 
by the long-distance behavior of the equations of motion. After rescaling $x=ref_{\pi}$, we obtain the ratio
$r/r_0=x/x_0$. The quantity $x_0$ has the meaning of a dimensionless size of a soliton 
(or rather in units of $(ef_{\pi})^{-1}$). 
The advantage of using (\ref{11}) is that all integrals involving
the profile function $F(x/x_0)$ can be evaluated analytically. 

The SU(3) extension of the Skyrme Lagrangian ${\cal L}$ uses set of parameters ${\hat x}, 
{\beta}', {\delta}'$ introduced in \cite{wei}:
\begin{eqnarray}
 {\hat x}&=&\frac{2m^2_K f^2_K}{m^2_{\pi}f^2_{\pi}} -1, \;\;\beta' =\frac{f^2_K - f^2_{\pi}}{4(1-{\hat x})}, 
\nonumber\\
\delta' &=& \frac{m^2_{\pi}f^2_{\pi}}{4} =\frac{m^2_K f^2_K}{2(1+{\hat x})}.
\label{34a}
\end{eqnarray}
The $\delta'$ term is required to split pseudoscalar meson masses, while the $\beta'$ term is required
to split pseudoscalar decay constants.

Including the previously introduced arctan ansatz for the profile function $F(r)$, 
we calculate the SU(3) extended classical soliton mass ${\cal E}_{\rm csol}$, 
the decuplet--octet mass splitting $\Delta$,
the antidecuplet--octet mass splitting $\overline\Delta$, i.e., 
the moment of inertia $\lambda_c$ for rotation in coordinate space, 
and the moment of inertia $\lambda_s$ for flavor rotations in the direction of the strange degrees of freedom,
except for the eighth direction \cite{wei,prat2}, 
and the symmetry breaking quantity $\gamma$. The quantity $\gamma$ is  
the coefficient in the SB 
piece ${\cal L}_{\rm SB} = -\frac{1}{2} \gamma (1-D_{88})$ of a total collective Lagrangian ${\cal L}$ 
and is linear in the SB parameter $(1-{\hat x})$. The above--mentioned quantities are given by
the following equations:
\begin{eqnarray}
{\cal E}_{\rm csol} 
&=& \hspace{-1mm}3{\sqrt 2}{\pi}^2\frac{f_{\pi}}{e} 
\label{27} \\
&\times& \left[x_0 + \frac{15}{16x_0} + \frac{2}{f^2_{\pi}}\left(3{\beta}'x_0 +
\frac{4}{3} \frac{\delta'}{e^2 f^2_{\pi}} x^3_0 \right)\right], \nonumber  \\
{\Delta} &=& \frac{3}{2\lambda_c (x_0)},\;\;
{\overline\Delta} =\frac{3}{2\lambda_s (x_0)} ,\label{28} \\
\lambda_c 
&=&\frac{\sqrt 2 \pi^2}{3e^3 f_{\pi}} \left[6\left( 1 + 
2\frac{\beta'}{f^2_{\pi}} \right)x^3_0 + \frac{25}{4} x_0\right],\label{29} \\
\lambda_s 
&=&\frac{\sqrt 2 \pi^2}{4e^3 f_{\pi}} \left[4\left( 1 - 
2(1+2{\hat x})\frac{\beta'}{f^2_{\pi}} \right)x^3_0 + \frac{9}{4} x_0\right],\label{30} \\
\gamma  
&=& 4 \sqrt 2 {\pi}^2 \frac{1-{\hat x}}{ef_{\pi}} \left({\beta}'x_0 -
\frac{4}{3} \frac{\delta'}{e^2 f^2_{\pi}} x^3_0 \right) .
\label{33} 
\end{eqnarray}
It is important to note that nowadays everybody agrees that the SU(3) extended Skyrme model 
classical soliton mass ${\cal E}_{\rm csol}$ receives to large value.
The consequence of this is unrealistic baryonic mass spectrum.
The ${\cal E}_{\rm csol}$ is connected with octet 
mass mean ${\cal M}_{\bf 8}$. From experiment we know 
${\cal M}_{\bf 8}=\frac{1}{8}\sum^8_{B=1}M^{\bf 8}_B =1151$ MeV.
Taking all that into account it is more appropriate to express mass formulas by 
${\cal M}_{\bf 8}$ instead by ${\cal E}_{\rm csol}$.
However, we are using the result for the classical soliton mass (\ref{27})
to obtain $x_0$, by minimalizing ${\cal E}_{\rm csol}$:
\begin{eqnarray}
x^2_0 = \frac{15}{8} \left[ 1 + \frac{6\beta'}{f^2_{\pi}} + 
\sqrt {\left(1+ \frac{6\beta'}{f^2_{\pi}}\right)^2 + 
\frac{30\delta'}{e^2 f^4_{\pi}}} \; \right]^{-1}.
\label{34} 
\end{eqnarray}
The dimensionless size of the skyrmion $x_0$  
includes dynamics of SB effects which takes place within skyrmion. It is clear from 
the above equation that a
skyrmion effectively shrinks when one ``switches on'' the SB effects and it shrinks more when the 
Skyrme charge $e$ receives smaller values.

To obtain the {\bf 8}, {\bf 10} and $\overline{\bf 10}$ absolute mass spectrum, we use the following 
definition of the mass formulas:
\begin{eqnarray}
M^{\bf 8}_B &=& {\cal M}_{\bf 8} - \frac{1}{2} \;\delta^{\bf 8}_B \;\gamma(x_0) , \nonumber \\
M^{\bf 10}_B &=& {\cal M}_{\bf 8} +\frac{3}{2\lambda_c (x_0)}- \frac{1}{2} \;\delta^{\bf 10}_B \;\gamma(x_0) , \nonumber \\
M^{\overline{\bf 10}}_B &=& {\cal M}_{\bf 8} +\frac{3}{2\lambda_s (x_0)}- \frac{1}{2} \;
\delta^{\overline{\bf 10}}_B \;\gamma(x_0) ,
\label{34b}
\end{eqnarray}
where ${\cal M}_{\bf 8}$ is defined earlier and the splitting constants $\delta^{\bf R}_B$ are given in 
Eqs. (17) to (19) of Ref. \cite{pra3}. Also, from 
experiment we know ${\cal M}_{\bf 10}=\frac{1}{10}\sum^{10}_{B=1}M^{\bf 10}_B 
=1382$ MeV.

Formulas (\ref{34b}) imply equal spacing for antidecuplets. From the existing experiments
($\Theta^+=1540$ MeV and $ \Xi^{--}_{3/2}=1861$ MeV) we estimate that spacing to be 
$\overline{\delta}=(1861-1540)/3=107$ MeV.
Next we estimate masses of antidecuplets N*=1647 MeV, $\Sigma^*_{\overline{10}} =1754$ MeV and the  
${\overline{\bf 10}}$ mean mass 
${\cal M}_{\overline{\bf 10}}=\frac{1}{10}\sum^{10}_{B=1}M^{\overline{\bf 10}}_B =1754$ MeV.
Finally we obtain the antidecuplet--octet mass splittings
${\overline\Delta}_{\rm exp} ={\cal M}_{\overline{\bf 10}}-{\cal M}_{\bf 8}=603$ MeV.
However, the decuplet--octet mass splittings ${\Delta}_{\rm exp} =231$ MeV represent the true experimental value.

Now we calculate the mass splittings $\Delta$ and $\overline\Delta$ for\\
(i) the SB with the approximation 
$f_{\pi} = f_K= 93$ MeV, ($\beta^{\prime} = 0, \;\;\delta^{\prime}=4.12 \times 10^7 \;{\rm MeV}^4$); and for\\ 
(ii) the SB with $f_{\pi} = 93$ MeV, $f_K = 113$ MeV,  
($\beta^{\prime}= -28.6 \;{\rm MeV}^2$ and $\delta^{\prime}=4.12 \times 10^7 \;{\rm MeV}^4$). \\
The results are presented in Table I.
\renewcommand{\arraystretch}{1.4}
\begin{table}[hbt]
\caption{The mass splittings $\Delta$ and $\overline\Delta$ for cases (i) and (ii) as a functions of $e$.}
\begin{center}
\begin{tabular}{|c|ccc|ccc|c|}
\hline
$ $ & $ $ & $ (\rm i) $ & $ $ & $ $ & $ (\rm{ii}) $ & $  $ & $ $\\
\hline
${\rm Mass \;Spl.}\backslash e$ & $3.4$ & $4.2$ & $4.6$ & $3.4$ & $4.2$ & $4.6$ & ${\rm Exp.}$  \\
\hline \hline
$ \Delta \; (\rm MeV)$ & $ 129 $ & $ 229$ & $ 294$ & $ 128$ & $ 227 $ & $ 291$ & $ 231 $  \\
\hline
$ \overline\Delta  \; (\rm MeV)$ & $ 354 $ & $ 621$ & $ 795$ & $ 273$ & $ 474 $ & $ 604$ & $ 603 $ \\
\hline \hline
\end{tabular}
\label{t:tab1.4}
\end{center}
\end{table}

We have chosen the three values of Skyrme charge $e=3.4;\;4.2;\;4.6$. The reason for this lies in the fact that 
in our minimal approach, case (ii), $e=3.4$ gives the best fit for 
the nucleon axial coupling constant $g_A = 1.25$ \cite{dppt}, 
$e=4.2$ fits nicely $\Delta_{\rm exp}$, and $e=4.6$ gives the best fit for $\overline\Delta_{\rm exp}$. 
However, from Table I, we see that a
certain middle value of $e(=4.2)$ supports also the case (i), i.e. in good agreement with experiment. 

The {\bf 8}, {\bf 10}, and $\overline{\bf 10}$ baryon mass spectrum (\ref{34b})
as a function of the SB effects and the Skyrme charge $e$ is given in Table II.
\renewcommand{\arraystretch}{1.2}
\begin{table}[hbt]
\caption{The {\bf 8}, {\bf 10}, and $\overline{\bf 10}$ baryon mass spectrum (MeV) 
for cases (i) and (ii) as a functions of $e$.}
\begin{center}
\begin{tabular}{|c|ccc|ccc|c|}
\hline
$ $ & $ $ & $ (\rm i) $ & $ $ & $ $ & $ (\rm{ii}) $ & $  $ & $ $\\
\hline
${\rm Mass}\backslash e$ & $3.4$ & $4.2$ & $4.6$ & $3.4$ & $4.2$ & $4.6$ & $
{\rm Exp}^{[34]}$  \\
\hline \hline
$ \rm N$ & $ 934 $ & $ 1024$ & $ 1051$ & $ 793$ & $ 934 $ & $ 977$ & $ 939 $  \\

$ \Lambda$ & $ 1079 $ & $ 1109$ & $ 1118$ & $ 1032$ & $ 1079 $ & $ 1093$ & $ 1116 $ \\

$ \Sigma$ & $ 1223 $ & $ 1193$ & $ 1184$ & $ 1270$ & $ 1223 $ & $ 1209$ & $ 1193 $  \\
 
$ \Xi$ & $ 1295 $ & $ 1236$ & $ 1218$ & $ 1390$ & $ 1295 $ & $ 1267$ & $ 1318 $  \\
 
$ \Delta$ & $ 1190 $ & $ 1327$ & $ 1403$ & $ 1130$ & $ 1287 $ & $ 1370$ & $ 1232 $  \\

$ \Sigma^*$ & $ 1280 $ & $ 1380$ & $ 1445$ & $ 1279$ & $ 1378 $ & $ 1442$ & $ 1385 $ \\

$ \Xi^*$ & $ 1371 $ & $ 1433$ & $ 1487$ & $ 1428$ & $ 1468 $ & $ 1514$ & $ 1530 $  \\

$ \Omega$ & $ 1461 $ & $ 1486$ & $ 1529$ & $ 1578$ & $ 1558 $ & $ 1587$ & $ 1672 $  \\

$ \Theta^+$ & $ 1325 $ & $ 1666$ & $ 1862$ & $ 1125$ & $ 1444 $ & $ 1611$ &
$1540^{[2, 3]} $  \\

$ \rm N^*$ & $ 1415 $ & $ 1719$ & $ 1904$ & $ 1274$ & $ 1535 $ & $ 1683$ & $ - $  \\
 
$ \Sigma^*_{\overline{10}}$ & $ 1505 $ & $ 1772$ & $ 1946$ & $ 1424$ & $ 1625 $ & $ 1755$ & $ - $ \\

$ \Xi^{--}_{3/2}$ & $ 1595 $ & $ 1825$ & $ 1988$ & $ 1573$ & $ 1715 $ & $ 1828$
& $ 1861^{[4]} $  \\
\hline \hline 
\end{tabular}
\label{t:tab1.4}
\end{center}
\end{table}
Since we are using the most simple version of the total Lagrangian (\ref{2a}),
i.e., we omit vector meson effects, the so-called static kaon fluctuations \cite{wei} and other fine--tuning effects 
in the expressions (\ref{27})--(\ref{34b}), our results given in Tables I and II,
do agree roughly with the other Skyrme model based estimates \cite{MP,pra3,dia1,wei1,WK,K}. 
In particular, our approach is similar to the one of Refs. \cite{WK,K}. The main difference is that our
Lagrangian is simpler, i.e. contains only SB proportional to $\lambda_8$, and that we are using the 
arctan ansatz approximation for the profile function $F(r)$. Comparing the pure Skyrme model
prediction of Ref. \cite{K} (fits A and B in Table 2) with our results for $e=4.2$, presented in Table II,
we have found up to the 8 \% differences. 
One of the reasons is due to the fact that the fits A and B in Table 2 of Ref. \cite{K} were obtained for 
different $e$'s, i.e. $e=3.96$ and $e=4.12$. Also, from our Table II one can see that for $e=4.2$, case (ii), 
mass spectrum differs from the experiment $\stackrel{<}{\sim} 8$\% for 
$\Omega^-$, $\Theta^+$ and $\Xi_{3/2}^{--}$. 
All other estimated masses are $\stackrel{<}{\sim}\; 5$\% different from experiment.
From Table II we conclude that in our minimal approach the best fit for 
{\bf 8}, {\bf 10}, and $\overline{\bf 10}$ baryon mass spectrum,
as a function of $e$ and for $f_{\pi}\not= f_K$, would lie between $e\simeq 4.2$ and $e\simeq 4.6$.

Symmetry breaking effects are generally very important and do improve theoretical estimates of the quantities 
like $\Delta$, $\overline\Delta$, the baryon mass spectrum etc. 
Our Tables I and II show implicitely that the inclusion of additional contributions, 
like vector meson contributions, 
the so-called static kaon fluctuations \cite{wei} and other fine--tuning effects into the SB Lagrangian \cite{WK}
does not change the results dramatically. On the contrary, the main effect is coming from the famous
$D_{88}$ term. The difference between $f_{\pi}$ and $f_K$ and the $e$ dependence are important.
All other contributions represent the fine tuning effects of the order of a few percent \cite{foot}. 
This is important for understanding the overall picture of the baryonic mass spectrum as well as for 
further study of other nonperturbative, higher-dimensional operator matrix elements in the Skyrme model \cite{dppt,tra}. 

\vspace{0.3cm}
We would like to thank T. Anti\v ci\' c and K. Kadija for  helpful discussions,
and M. Prasza{\l}owicz for careful reading of the manuscript.
This work was supported by the Ministry of Science and Technology of the Republic of Croatia under Contract 0098002.

\end{document}